\documentclass[conference]{IEEEtran}
\IEEEoverridecommandlockouts
\usepackage{cite}
\usepackage{amsmath,amssymb,amsfonts}
\usepackage{algorithmic}
\usepackage{soul,color}
\usepackage{graphicx}
\usepackage{textcomp}
\usepackage{xcolor}
\usepackage{multirow}
\usepackage{url}
\def\BibTeX{{\rm B\kern-.05em{\sc i\kern-.025em b}\kern-.08em
    T\kern-.1667em\lower.7ex\hbox{E}\kern-.125emX}}
\begin{document}

\title{Streaming Remote rendering services: \\Comparison of QUIC-based and WebRTC Protocols \\
}

\author{
\IEEEauthorblockN{Daniel Mejías\IEEEauthorrefmark{1}, Inhar Yeregui\IEEEauthorrefmark{1}, Ángel Martín, Roberto Viola}
\IEEEauthorblockA{Fundación Vicomtech\\
Basque Research and Technology Alliance\\
San Sebastián, 2009 Spain\\
Email: damejias@vicomtech.org}
\IEEEauthorblockA{\IEEEauthorrefmark{1}PhD Candidate at UPV/EHU}
\and
\IEEEauthorblockN{Pablo Angueira, Jon Montalb\'an}
\IEEEauthorblockA{Department of Communications\\Engineering\\ 
University of the Basque Country\\
Bilbao, 48013 Spain\\
Email: \{pablo.angueira, jon.montalban\}@ehu.eus}
\and
}
\maketitle

\IEEEoverridecommandlockouts
\IEEEpubid{\begin{minipage}{\textwidth}\ \\\\\\\\\\[12pt]\centering
979-8-3315-1998-8/25/\$31.00 ©2025 IEEE
\end{minipage}}

\begin{abstract}
The proliferation of Extended Reality (XR) applications, requiring high-quality, low-latency media streaming, has driven the demand for efficient remote rendering solutions. 
This paper focuses on holographic conferencing in virtual environments and their required uplink and downlink media transmission capabilities.
By examining Media over QUIC (MoQ), Real-time Transport Protocol (RTP) over QUIC (RoQ), and Web Real-Time Communication (WebRTC), we assess their latency performance over Wi-Fi and 5G networks.
Improvements of approximately 30\% in latency and 60\% in connection startup are expected in QUIC-based protocols compared to WebRTC.
The experimental setup transmits a remote-rendered virtual experience using real-time video streaming protocols to provide the content to the participant.
Our findings contribute to understanding the maturity of streaming protocols, particularly within open-source frameworks, and evaluate their suitability in supporting latency-sensitive XR applications. The study highlights specific protocol advantages across varied remote rendering scenarios, informing the design of future XR communication solutions.
\end{abstract}


\section{Introduction}

The demand for remote rendering systems is increasing. Modern Extended Reality (XR) applications for gaming and virtualised environments have become highly demanding to meet the needs across various fields; however, hardware requirements have also increased significantly\cite{diepstraten2004remote}. Developing environments for assistance, education, and entertainment now demands detailed scenarios and remote accessibility for users anywhere\cite{li2010augmented}. This requires high hardware capacities and low-latency streaming. For this reason, remote rendering systems are essential, enabling access to high-performance hardware that may not be available to the user locally.

Protocols like Web Real-Time Communication (WebRTC) are currently used in XR to enable low-latency, real-time communication \cite{rfc8825}. However, with the advent of the new Hypertext Transfer Protocol version 3 (HTTP/3) \cite{rfc9114}, the use of the Quick User Datagram Protocol (UDP) Internet Connections (QUIC) protocol is being proposed as an alternative \cite{rfc9369}.
Based on UDP, QUIC simplifies connection negotiation and data transport while introducing novel capabilities such as connection migration across networks, integrated end-to-end encryption, stream multiplexing, and multi-path delivery\cite{cook2017quic}. It also features more efficient congestion management, making it a robust choice for modern Internet applications. QUIC offers significant advancements over traditional protocols like TCP (Transmission Control Protocol) and UDP, positioning it as a promising solution for enhancing the performance and flexibility of XR applications. Its design aims to surpass WebRTC in both capacity and impact.

XR applications are particularly challenging due to their stringent performance requirements regarding latency, throughput usage, and tolerance to bytes loss. Therefore, studying the most suitable protocol and communication stack is both necessary and highly relevant. To test the impact of each analyzed option, this work deploys a rendering service implementation
on the edge infrastructure of an experimental and private 5G Standalone (SA) network. 
This service delivers a video stream that includes a personalized scenario where the user is a passive element, using the communication technologies under study. The main contributions of this paper are:
\begin{itemize}
    \item Comparative Analysis of Protocols for XR Applications. This study conducts a comprehensive latency comparison of different protocols—Media over QUIC (MoQ), Real-Time Protocol (RTP) over QUIC (RoQ), and WebRTC—when integrated with remote rendering services. 
    \item Evaluation across heterogeneous networks. The protocols are evaluated under both Wi-Fi and private 5G networks, setting expectations on their performance across typical network environments for remote rendering.
    \item Open-source Framework Maturity. The paper assesses the maturity and robustness of an open-source framework supporting each protocol, addressing practical considerations when streaming media of remote rendering services.
\end{itemize}

This paper is structured as follows. Section \ref{sec:related} covers the state of the art of real-time protocols for remote XR environments. Then, Sections \ref{sec:implemented} and \ref{sec:results} explain the implementation of the remote rendering service and analyze the results obtained with it for different protocols and cases. Finally, in Section \ref{sec:conclusion} we present our conclusions and recommendations for future research. 

\section{Related Work}
\label{sec:related}


\subsection{Remote Rendering Platforms}

Remote rendering is a technique that involves processing and rendering 2D or 3D graphics on a remote server. This technology is particularly relevant for XR applications, as it enables devices with limited resources to access and utilize advanced graphical capabilities. In recent years, the evolution of video games, entertainment, and education has increasingly shifted towards remote environments, allowing users to experience services seamlessly from any location. 

In the recent literature, several approaches to remote rendering have been proposed, many of which are closely related to cloud gaming and focus on achieving Ultra-Low Latency (ULL) in interactive communication
\cite{bassbouss2023metaverse}, \cite{corry2023evaluating}. One notable example is CloudXR \footnote{https://developer.nvidia.com/cloudxr-sdk}, a solution developed by NVIDIA that enables communication with rendering engines like Unity or Unreal Engine through a proprietary communication channel. However, this solution has certain limitations, including compatibility restricted to specific devices with paid applications and the requirement of a licensing agreement for its use. Other remote rendering implementations incorporate integration with WebRTC \cite{oros2020renderlink}. This low-latency communication protocol is an excellent option, as it supports multi-user communication and enables the transmission of video, audio, and interactive data with end-to-end connectivity \cite{yeregui2024edge}.
Although new protocols for low-latency communication are emerging, there is a need for a system that ensures open compatibility with various protocols.

\subsection{Streaming protocols for XR}
In XR streaming environments, real-time transmission is essential. Technologies like WebRTC enable real-time streaming through a bidirectional channel, supporting audio, video, and interactivity. WebRTC relies on the RTP protocol at the application layer to achieve peer-to-peer video transmission, making it a reliable solution for real-time communication \cite{10733608}. However, there is an increasing demand for protocols that offer not only low-latency performance but also enhanced security and greater flexibility within the network.
QUIC introduces significant advancements in security, flexibility, and performance, particularly for low-latency scenarios. Built on the UDP transport layer, QUIC was originally proposed by Google and has since then been standardized. Leveraging UDP at the transport layer preserves standardization while facilitating new integrations at the user level. This architecture enables QUIC to implement flow multiplexing, treating multiple requests from the same client as a single UDP flow between the client and server\cite{cook2017quic}.
Real-time streaming in QUIC leverages its ability to transmit encoded video and audio data efficiently as streams\cite{rfc9000} or datagrams \cite{rfc9221}. Among the recent advancements, RTP over QUIC (RoQ) extension stands out, enabling the seamless transmission of RTP packets over the QUIC protocol \cite{ietf-avtcore-rtp-over-quic-12}. This integration combines the low-latency and multiplexing benefits of QUIC with the real-time capabilities of RTP, making it a robust solution for modern streaming applications. 
Another QUIC extension is Media over QUIC (MoQ), which aims to provide a simple, low-latency solution for media ingestion and distribution \cite{ietf-moq-transport-07}. MoQ facilitates publisher/subscriber communication and employs a network relay deployment to buffer and transmit fragments of data efficiently, ensuring low-latency delivery.

\subsection{XR over wireless networks}

Efficient support for XR services presents new challenges for current and future wireless networks. 5G network deployments are particularly well-suited to meet these demands, offering high performance and low latency. Furthermore, the 3rd Generation Partnership Project (3GPP), in its Release 18, defines the minimum performance indicators required to support XR services effectively\cite{10296869}, \cite{10666455}.

A similar challenge has emerged within the Wi-Fi standard, prompting efforts to achieve deterministic wireless connectivity with ultra-low latency. Wi-Fi 6 establishes the baseline requirements for efficient communication in XR services and low latency \cite{10520633}. However, new developments are underway to incorporate time-sensitive networking (TSN) capabilities into Wi-Fi, aiming to meet even more stringent performance demands \cite{10034532}.

A key feature of QUIC is its support for multipath communication, which allows the simultaneous use of multiple network paths. This capability enables QUIC to leverage different interfaces, such as Wi-Fi and 5G, to optimize both connection speed and stability. By dynamically adapting to varying network conditions, QUIC enhances the overall user experience, particularly in mobile environments where connectivity can be unpredictable \cite{de2017multipath}.




In this research, we propose the development of a solution based on the Unity rendering engine, designed for deployment in a distributed containerized system. At the client level, it can be integrated into mobile applications and web services. The focus is on the development of the Remote Renderer, leveraging low-latency communication protocols, particularly those based on QUIC.

\begin{figure*}[hbt!]
\centering
\includegraphics[width=1.0\textwidth]{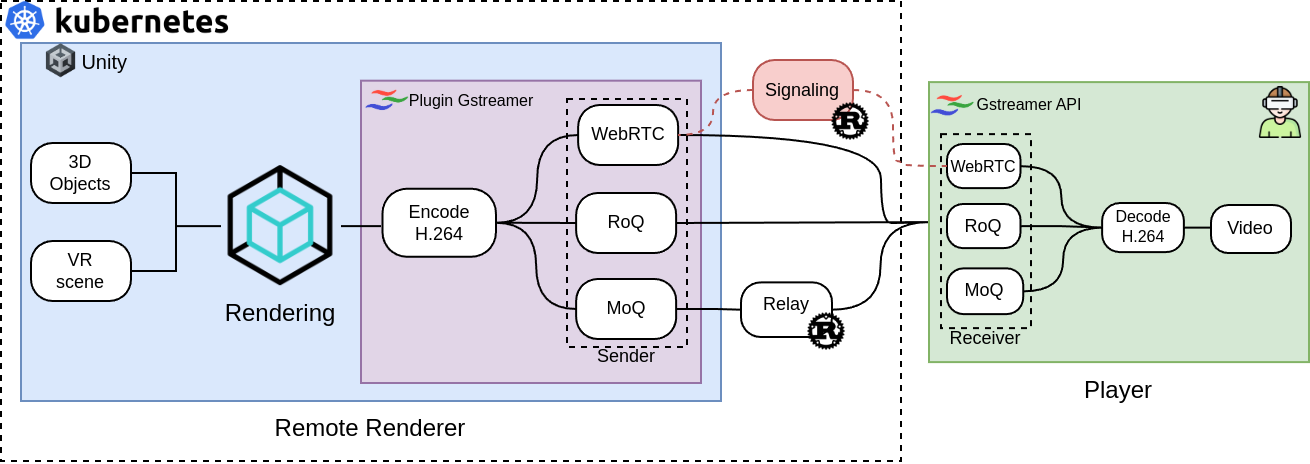}
\caption{Architecture for implementing communication between the remote renderer and the player}
\label{fig:architecture}
\end{figure*}

\section{Implementation of a Remote Rendering Service}
\label{sec:implemented}
\subsection{Use Case}

In a holoconferencing system with two or more participants, real-time communication is critical. A potential use case is a live event, such as a concert or show, where participants act as spectators or passive clients. The system consists of a Remote Renderer deployed in a distributed containerized environment and a player, which can be a laptop. In this scenario, the immersive XR experience requires the lowest possible communication latency. Figure \ref{fig:architecture} illustrates the proposed communication between the remote renderer and the player for each of the analyzed protocols.

The Remote Renderer consists of two main software components: the Unity application, developed in C\#, and the native rendering plugin, which uses Gstreamer libraries and is developed in C++ and dynamically linked to the main application. 

The main Unity application acts as the backbone, coordinating the interactions between the different components. It is based on the Unity rendering engine. It uses the power of NVIDIA GPUs to process the heterogeneous content embedded in a VR scene, including audio and volumetric video, and generate rendered raw audio and 360$^{\circ}$ video streams.

The native rendering plugin\footnote{https://docs.unity3d.com/Manual/NativePlugins.html}
is a custom Unity plugin that serves as a bridge between the main Unity application and the multimedia framework responsible for creating a protocol-specific multimedia pipeline. For this development, we implemented the GStreamer framework, which provides encoding and communication functionalities for multiple protocols through its libraries.


\subsection{Suitable Protocols}
 
GStreamer is integrated by invoking its native Application Programming Interfaces (APIs) to construct and manage the media processing pipeline. The GStreamer element \textit{appsrc} \cite{gstappsrc} enables the injection of rendered video buffers into GStreamer pipeline, which are subsequently encoded using NVIDIA hardware encoders. The video is consistently encoded with the H.264 codec. The final operation within the pipeline varies depending on the protocol. In this work, we study  the currently supported protocols and their corresponding operations.

    WebRTC, which is Based on RTP,  begins with an initial negotiation for peer-to-peer communication through a signaling server. The signaling server is developed in RUST\footnote{https://www.rust-lang.org/es} and is an official implementation for webRTC elements in Gstreamer. Once established, it creates a communication channel where the encoded streams are encapsulated within a WebRTC stream. 
    
    The player is developed in RUST and is responsible for receiving the encapsulated RTP stream and decoding the data. NVIDIA hardware decoding is used for decoding.
    
    In RoQ, the communication negotiation is established in a peer-to-peer mode, after which the encoded video is encapsulated into RTP streams with adaptations for the QUIC protocol. GStreamer officially supports this development; however, it is still in an early stage and lacks full stability. 
    
    Similar to WebRTC, the player is developed in RUST and is responsible for receiving the stream, decapsulating the RTP packets, and decoding the data  using NVIDIA hardware decoding.
     
     In the case of MoQ, a fragmented MP4 (fMP4) muxer is used to encapsulate the encoded stream. Then, an unofficial GStreamer element implements MoQ to transmit the fMP4 stream. 
    In this case, a relay-based communication model is employed, where the server pushes the data, and the client requests it. The relay server is developed in RUST. Since these elements are not included in the stable, officially maintained versions of GStreamer, they are less mature than other alternatives and currently do not support audio.
    
    Similar to the remote renderer, the player utilizes an unofficial GStreamer element to receive the fMP4 stream, which is then decoded using GStreamer elements \cite{curley2024media} and NVIDIA hardware decoding.

\label{sec:results}

\begin{figure}[!t]
\centering
\includegraphics[width=0.48\textwidth]{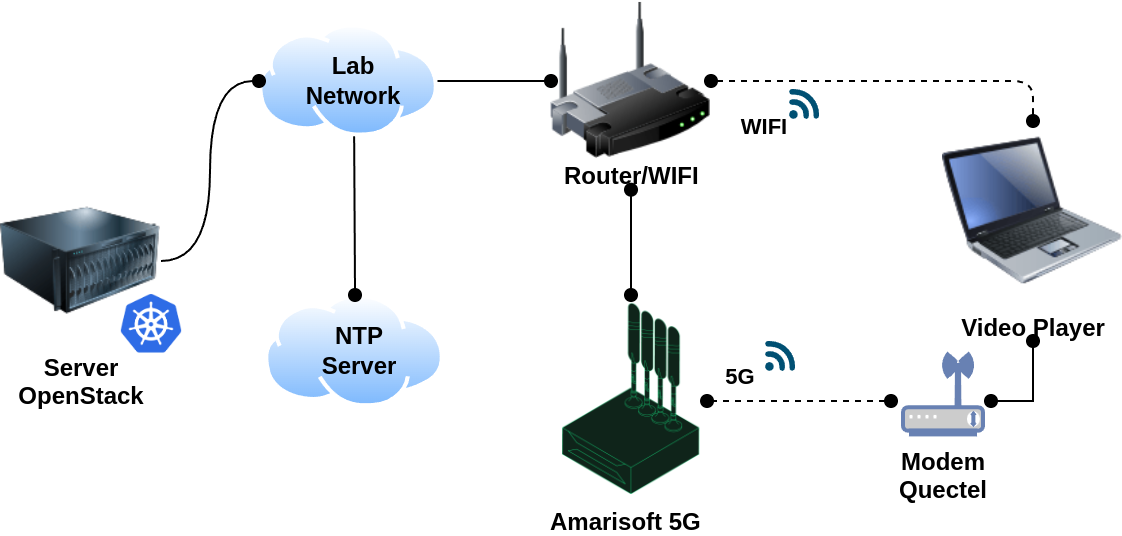}
\caption{Testbed for Wi-Fi and 5G latency measurement.}
\label{fig:testbed}
\end{figure}

\subsection{Testing Setup}

Figure \ref{fig:testbed} illustrates the testbed configuration, which includes 5G and Wi-Fi for communication between the remote renderer and the player. 

    The server consists of the remote renderer, the WebRTC signaling server, and the MoQ relay server. The deployment was carried out in a Kubernetes\footnote{https://kubernetes.io/} cluster, where each service was allocated to separate pods. The Kubernetes deployment was carried out using a Helm chart\footnote{https://helm.sh/}. The Kubernetes cluster is
    hosted on a Virtual Machine (VM) equipped with 8 Central Processing Unit (CPU) cores, 32 GB of Random Access Memory (RAM), and an NVIDIA L4 virtual Graphics Processing Unit (vGPU) with 8 GB of dedicated memory (vGPU Profile 8Q\footnote{\url{https://docs.nvidia.com/vgpu/sizing/virtual-workstation/latest/methodology.html#vgpu-profiles}}).
    
    The Lab Network is an internal network where OpenStack servers are deployed, designed for testing and internet access.
    
    A Wi-Fi 6-compatible router operating on the 5 GHz frequency band was used to facilitate communication between the remote rendering server and the video playback device. Alternatively, it can redirect traffic to a 5G network instead of Wi-Fi. The Wi-Fi configuration is Access Point.
    
   For the 5G network, we used the AMARI Callbox Mini \cite{AMARI}, a 5G Core and Radio Access Network (RAN) device that implements the 3rd Generation Partnership Project (3GPP) Release 15 \cite{rel15}. It is configured to use the N78 band, 20 MHz bandwidth, 30 MHz subcarrier spacing, max modulation of 256 Quadrature Amplitude Modulation (QAM), 2 antennas for downlink and 1 for uplink.

    As a 5G modem, the Quectel RM500Q was used, which implements 3GPP Release 15 and is connected to the video player via USB.
     
      A laptop was used where the video stream is rendered through a dedicated video player application. Any standard laptop with basic specifications is sufficient, provided it includes a built-in Wi-Fi 6 adapter and the necessary drivers for the Quectel modem. he tests were conducted on a laptop equipped with an AMD Ryzen 5800H CPU, 16 GB of Double Data Rate 4 (DDR4) RAM, and an NVIDIA GTX 1650 graphics processing unit (GPU).
    
    For synchronization, the router grants internet access, allowing both the remote rendering server and the video playback device to act as NTP clients and synchronize with a public Network Time Protocol (NTP) server, ensuring accurate latency measurements.

During the tests, the \textit{tshark}\footnote{https://tshark.dev/} tool is used to capture and analyze the network traffic, with the server acting as the sender and the player as the receiver. The traffic generated by both parties is stored in a packet capture (PCAP) file for subsequent comparison and analysis.

\section{Results}

For the comparison of the WebRTC, MoQ, and RoQ protocols, two tests were designed, involving network packet captures and image captures to measure transmission and reception times.
\subsection{Latency and Resource Test}
The first test measures the average 
 time required to establish communication between the client and server (startup time)
and End-to-End latency over a Wi-Fi 6 and 5G network.
The rendered video is encoded using the H.264 codec, comparing three configurations: 1080p at 15 Mbps, 720p at 10 Mbps, and 480p at 5 Mbps. In all configurations, the video is rendered at 30 FPS, encoded with a Gruops of Picture (GOP) size of 5 and main profile.
Additionally, resource usage in the Remote Renderer was measured in terms of CPU and GPU utilization during encoding, while in the player, CPU and GPU usage were analyzed during decoding.

Table \ref{tab:RTP_Comp} presents the results of this first test, highlighting the following key observations.
The startup time measurements indicate that QUIC-based protocols outperform WebRTC, primarily because WebRTC requires a negotiation process to establish the communication path, whereas QUIC protocols only need to perform a UDP handshake. Similarly, MoQ achieves better startup performance than RoQ. This is because, after completing the handshake, RoQ must wait for the transmission to begin, whereas MoQ already generates traffic to the relay and only needs to request a redirection of the relay server's traffic. Finally, encoding parameters do not impact startup time, as the measurement only considers the time elapsed until the first frame is received.
Regarding network differences, the Wi-Fi network demonstrates better startup times compared to 5G, with an approximate improvement of 200 milliseconds. 
This difference may stem from the fact that 5G requires routing through the core, whereas Wi-Fi operates as a direct access point.
In the latency comparison, RoQ achieves the best performance among the three protocols, with an improvement of 90 milliseconds over WebRTC. However, MoQ also exhibits the highest latency values due to the use of a relay for communication between the server and the player, which increases latency by approximately 100\%. In the encoding parameters, it is observed that the configuration with the lowest bitrate and resolution results in the lowest latency, while latency increases as the encoding parameters (bitrate and resolution) are increased. Regarding network differences, the 5G network outperforms Wi-Fi, with a variable latency reduction ranging from 30 to 100 milliseconds.
In this case, since communication is already established, routing does not impact latency, resulting in improved latency values. Additionally, both networks exhibit good performance due to the absence of competing traffic.

Regarding CPU and GPU usage of the Remote Renderer and the player, the results can be seen in Table \ref{tab:usage}. 
CPU consumption during encoding exceeds 100\%, as the server utilizes multiple cores for processing. The CPU measurements represent the total percentage of usage across rendering, processing, and transmission by the protocol.
The highest CPU consumption is in rendering.
Another important observation is that CPU usage increases as the video parameters (bitrate and resolution) increase. This is because higher rendering and processing parameters generate more data, requiring additional computational resources for processing. Finally, when comparing protocols, WebRTC exhibits the lowest CPU consumption. This can be attributed to their maturity and the optimization of existing libraries for efficient resource utilization, in contrast to MoQ and RoQ, which are still in earlier stages of development.

GPU consumption during encoding increases in accordance with the encoding parameters. This is due to the higher data volume being processed as bitrate and resolution increase. Notably, in the 480p/5Mbps configuration, the GPU usage is reported as 0\%. This occurs because the actual usage is below 1\%, and the measurement system lacks the granularity to display more precise values. Regarding the differences between protocols, all exhibit the same GPU consumption. Since the encoding parameters remain identical across protocols, the expected GPU usage remains consistent.

Regarding resource consumption during decoding on the player, it is observed that higher encoding parameters (bitrate and resolution) lead to increased resource usage. In terms of GPU utilization, all protocols exhibit similar consumption, as the video stream remains the same across them. However, in terms of CPU usage, WebRTC consumes the most resources due to the higher complexity and additional elements required in GStreamer for its reception. RoQ, on the other hand, is the least resource-intensive due to its simpler player implementation. However, it is also more error-prone, as it lacks mechanisms for handling transmission errors, indicating that its implementation is not yet mature.


\subsection{Network Performance Test}
In the second test, each protocol is analyzed, evaluating throughput usage, jitter, and bytes loss at the player using the PCAP file obtained with the \textit{tshark} tool. Network traffic is captured for two minutes, and the test is performed over both 5G and Wi-Fi networks. In this case, the video rendered by the remote renderer is encoded using H.264 at 1080p, 30 FPS, and a bitrate of 15 Mbps.

Table \ref{tab:traffic} presents the results of the second test. Some of these results indicate that QUIC libraries and implementations are still in an early stage of development, exhibiting limited maturity.
The throughput measurements aim to assess the protocol's overhead efficiency. The results indicate that RoQ achieves the lowest data transmission values, followed by WebRTC, suggesting that these protocols have lower overhead. In contrast, MoQ exhibits the highest data transmission values, likely due to intrinsic protocol overhead. Regarding the differences between 5G and Wi-Fi networks, no significant variations in throughput were observed.


In terms of jitter, WebRTC demonstrates the most stable performance across both networks, suggesting it has more robust packet arrival handling and better latency jitter compensation. In contrast, RoQ and MoQ exhibit higher jitter over Wi-Fi, although their performance improves on the 5G network. This suggests that these protocols may still lack maturity in their packet handling.

Finally, regarding byte loss, WebRTC demonstrates performance comparable to or better than MoQ and RoQ. In the tests conducted with both QUIC-based protocols, higher byte loss was observed in Wi-Fi environments. In contrast, the 5G network showed improved performance, with values similar to those of WebRTC.

\renewcommand{\arraystretch}{1.2}
\begin{table}
    \centering
\setlength\tabcolsep{2.5pt} 
\caption{Comparison of protocols with QUIC vs RTP for video at 30FPS over 5G and Wi-Fi network}
    \begin{tabular}{|c|c|c|c|} \hline 
         \textbf{Encoding}&\textbf{Protocol}&\textbf{Startup} & \textbf{Latency} \\
          \textbf{bitrate}& &\textbf{(ms)}& \textbf{(ms)}\\  \hline 
         \multicolumn{4}{|c|}{\textbf{Wi-Fi}}\\ \hline
         \multirow{3}{*}{1080p/15Mbps}&WebRTC&1421.0& 288.83\\  
         \cline{2-4} 
         &RoQ&948.4& 215.00\\  
         \cline{2-4} 
         &MoQ&532.2& 559.83\\ \hline 
         \multirow{3}{*}{720p/10Mbps}&WebRTC&1451.2& 259.92\\ 
         \cline{2-4} 
         &RoQ&877.8& 189.83\\ 
         \cline{2-4} 
         &MoQ&525& 537.25\\ \hline
            \multirow{3}{*}{480p/5Mbps}& WebRTC&1498.6&  249.08\\
            \cline{2-4} 
            & RoQ&797.4&  167.83\\
            \cline{2-4} 
            & MoQ&495.2&  431.67\\ \hline
        \multicolumn{4}{|c|}{\textbf{5G}}\\ \hline
         \multirow{3}{*}{1080p/15Mbps}&WebRTC&1635.4& 269.92\\ 
         \cline{2-4} 
         &RoQ&1101& 125.67\\ 
         \cline{2-4} 
         &MoQ&810.4& 538.00\\ \hline 
         \multirow{3}{*}{720p/10Mbps}&WebRTC&1703.2& 253.08\\ 
         \cline{2-4} 
         &RoQ&988.4& 122.17\\ 
         \cline{2-4} 
         &MoQ&670.8& 524.08\\ \hline
            \multirow{3}{*}{480p/5Mbps}& WebRTC&1567.4&  234.08\\ 
            \cline{2-4} 
            & RoQ&892.4&  129.25\\ 
            \cline{2-4} 
            & MoQ&626.2&  533.92\\ \hline
    \end{tabular}

\label{tab:RTP_Comp}
\end{table}

\renewcommand{\arraystretch}{1.2}
\begin{table}[]
\caption{Comparison of CPU and GPU usage protocols in Remote Renderer vs Player for video at 30 FPS}
\label{tab:usage}
\begin{tabular}{cl|cc|cc|}
\cline{3-6}
 & \multicolumn{1}{c|}{} & \multicolumn{2}{c|}{\textbf{Remote Renderer}} & \multicolumn{2}{c|}{\textbf{Player}} \\ \hline
\multicolumn{1}{|c|}{\textbf{Encoding}} & \multicolumn{1}{c|}{\multirow{2}{*}{\textbf{Protocol}}} & \multicolumn{1}{c|}{\textbf{CPU}} & \textbf{GPU ENC} & \multicolumn{1}{c|}{\textbf{CPU}} & \textbf{GPU DEC} \\
\multicolumn{1}{|c|}{\textbf{bitrate}} & \multicolumn{1}{c|}{} & \multicolumn{1}{c|}{\textbf{(\%)}} & \textbf{(\%)} & \multicolumn{1}{c|}{\textbf{(\%)}} & \textbf{(\%)} \\ \hline
\multicolumn{1}{|c|}{\multirow{3}{*}{\begin{tabular}[c]{@{}c@{}}1080p\\ 15Mbps\end{tabular}}} & WebRTC & \multicolumn{1}{c|}{275} & \multicolumn{1}{c|}{2} & \multicolumn{1}{c|}{17} & 5 \\ \cline{2-6} 
\multicolumn{1}{|c|}{} & RoQ & \multicolumn{1}{c|}{305} & \multicolumn{1}{c|}{2}  & \multicolumn{1}{c|}{5} & 5 \\ \cline{2-6} 
\multicolumn{1}{|c|}{} & MoQ & \multicolumn{1}{c|}{303} & \multicolumn{1}{c|}{2}  & \multicolumn{1}{c|}{12} & 6 \\ \hline
\multicolumn{1}{|c|}{\multirow{3}{*}{\begin{tabular}[c]{@{}c@{}}720p\\ 10Mbps\end{tabular}}} & WebRTC & \multicolumn{1}{c|}{257} & \multicolumn{1}{c|}{1} & \multicolumn{1}{c|}{10} & 2 \\ \cline{2-6} 
\multicolumn{1}{|c|}{} & RoQ & \multicolumn{1}{c|}{300} & \multicolumn{1}{c|}{1} & \multicolumn{1}{c|}{3} & 3 \\ \cline{2-6} 
\multicolumn{1}{|c|}{} & MoQ & \multicolumn{1}{c|}{316} & \multicolumn{1}{c|}{1} & \multicolumn{1}{c|}{8} & 3 \\ \hline
\multicolumn{1}{|c|}{\multirow{3}{*}{\begin{tabular}[c]{@{}c@{}}480p\\ /5Mbps\end{tabular}}} & WebRTC & \multicolumn{1}{c|}{150} & \multicolumn{1}{c|}{0} & \multicolumn{1}{c|}{7} & 1 \\ \cline{2-6} 
\multicolumn{1}{|c|}{} & RoQ & \multicolumn{1}{c|}{247} &  \multicolumn{1}{c|}{0} & \multicolumn{1}{c|}{3} & 1 \\ \cline{2-6} 
\multicolumn{1}{|c|}{} & MoQ & \multicolumn{1}{c|}{260} &  \multicolumn{1}{c|}{0} & \multicolumn{1}{c|}{6} & 1 \\ \hline
\end{tabular}
\end{table}

\begin{table}
    \centering
\caption{Traffic Comparison between QUIC vs RTP Protocols  for 1080p/15Mbps/30FPS video over 5G and Wi-Fi network}
\label{tab:traffic}
    \begin{tabular}{|c|c|c|c|} \hline 
         \textbf{Protocol}& \textbf{Throughput}& \textbf{Jitter}&\textbf{Bytes}\\ 
 &\textbf{usage (kbps)}&\textbf{(ms)}&\textbf{Loss (\%)}\\ \hline 
 \multicolumn{4}{|c|}{\textbf{Wi-Fi}}\\ \hline
 WebRTC& 13782& 0.66&0.02\\ \hline 
 RoQ& 11002& 4.91&2.67 \\ \hline
 MoQ& 15920& 3.55& 2.85\\  \hline
  \multicolumn{4}{|c|}{\textbf{5G}}\\ \hline
 WebRTC& 13783& 0.64&1.15\\ \hline 
 RoQ& 11304& 0.87&1.00\\ \hline
 MoQ& 15988& 0.57&1.00\\ \hline
    \end{tabular}

\end{table}


\section{Conclusions And Future Work}
\label{sec:conclusion}

This paper presents a comprehensive analysis of the MoQ, RoQ, and WebRTC protocols, focusing on their latency performance when integrated into remote rendering services for dedicated XR applications in Wi-Fi and 5G network environments. The analysis underscores the importance of high-quality, low-latency transmission, particularly for applications requiring interactive audiovisual content and volumetric data.
Additionally, the proposed solution is deployed in a distributed Kubernetes environment, facilitating future development and experimentation with cloud-based infrastructures.


The results show that QUIC-based protocols, like RoQ, have the potential to outperform WebRTC in low-latency transmission, with RoQ achieving the fastest connection startup but suffering from higher overall latency due to its communication mechanism. 
Additionally, an evaluation of the Open-source framework maturity of the QUIC framework and libraries was carried out, highlighting that, while WebRTC is highly optimized, QUIC is still at an early stage of development and evolution towards standardization.

For future research, we recommend extending the study to explore the performance of QUIC multipath when handling multiple concurrent streams, including video, audio, and data. Furthermore, investigating the feasibility and impact of bidirectional communication would provide valuable insights into enhancing interactive remote rendering applications.


\section*{Acknowledgment}

This research was supported by the SNS-JU Horizon Europe Research and Innovation programme, under Grant Agreement 101096838 for 6G-XR project.

\bibliographystyle{IEEEtran}
\bibliography{IEEEabrv,main.bib}


\end{document}